\documentclass[trackchanges]{aastex63}
\hypersetup{linkcolor=red,citecolor=blue,filecolor=cyan,urlcolor=magenta}

\received{}
\revised{}
\accepted{}
\submitjournal{PASP}

\shorttitle{Gamma-ray location of FSRQs}
\shortauthors{Huang et al.}

\begin{document}

\title{Constraining the $\gamma$-ray Emission Region for Fermi-Detected FSRQs by the Seed Photon Approach}

\correspondingauthor{Zhiyuan Pei}
\email{peizy@gzhu.edu.cn}
\correspondingauthor{Junhui Fan}
\email{fjh@gzhu.edu.cn}

\author{Danyi Huang}
\affiliation{School of Physics and Materials Science, Guangzhou University, Guangzhou 510006, People's Republic of China}

\author{Ziyan Li}
\affiliation{School of Physics and Materials Science, Guangzhou University, Guangzhou 510006, People's Republic of China}

\author{Jiru Liao}
\affiliation{School of Physics and Materials Science, Guangzhou University, Guangzhou 510006, People's Republic of China}

\author{Xiulin Huang}
\affiliation{School of Physics and Materials Science, Guangzhou University, Guangzhou 510006, People's Republic of China}

\author{Chengfeng Li}
\affiliation{School of Physics and Materials Science, Guangzhou University, Guangzhou 510006, People's Republic of China}

\author{Yanjun Qian}
\affiliation{School of Physics and Materials Science, Guangzhou University, Guangzhou 510006, People's Republic of China}

\author[0000-0002-4970-3108]{Zhiyuan Pei}
\affiliation{School of Physics and Materials Science, Guangzhou University, Guangzhou 510006, People's Republic of China}
\affiliation{Center for Astrophysics, Guangzhou University, Guangzhou 510006, People's Republic of China}
\affiliation{Astronomy Science and Technology Research Laboratory of Department of Education of Guangdong Province, Guangzhou 510006, People's Republic of China}
\affiliation{Key Laboratory for Astronomical Observation and Technology of Guangzhou, Guangzhou 510006, People's Republic of China}

\author[0000-0002-5929-0968]{Junhui Fan}
\affiliation{School of Physics and Materials Science, Guangzhou University, Guangzhou 510006, People's Republic of China}
\affiliation{Center for Astrophysics, Guangzhou University, Guangzhou 510006, People's Republic of China}
\affiliation{Astronomy Science and Technology Research Laboratory of Department of Education of Guangdong Province, Guangzhou 510006, People's Republic of China}
\affiliation{Key Laboratory for Astronomical Observation and Technology of Guangzhou, Guangzhou 510006, People's Republic of China}

\begin{abstract}
The location of $\gamma$-ray emitting region in blazars has been an open issue for several decades and is still being debated. We use the Paliya et al. sample of 619 $\gamma$-ray-loud flat-spectrum radio quasars with the available spectral energy distributions, and employ a seed photon factor approach, to locate the $\gamma$-rays production region. This method efficiently set up a relation between the peak frequencies and luminosities for the synchrotron emission and inverse Compton scattering, together with a combination of the energy density and characteristic energy for the external seed photon field, namely, $\sqrt{U_0}/\epsilon_0$, an indicative factor of seed photons (SF) in units of Gauss. By means of comparing it with canonical values of broad-line region and molecular dusty torus, we principally ascertain that the GeV emission is originated far beyond the BLR and close to the DT---farther out at pc scales from the central black hole, which supports a {\it far-site} scenario for $\gamma$-ray blazars. We probe the idea that inverse Compton scattering of infrared seed photons is happening in the Thomson regime. This approach and our findings are based on the validity of the External Compton model, which is applicable to understand the GeV emission mechanism in FSRQs. However, the completeness of this framework has been challenged by reports of neutrino emission from blazars. Thus we also shed new light on the neutrino production region by using our derived results since blazars are promising neutrino emitters.
\end{abstract}


\section{Introduction}

As the most distinctive subclass of active galactic nuclei (AGN), blazars are characterized by high-amplitude rapid variability, apparent superluminal motion in their pc-scale jet, core-dominated nonthermal continuum, and strong emission over the entire electromagnetic spectrum, since their relativistic jets are oriented very close to the observer's line-of-sight \citep{Wil92, UP95, Rom02, Fan05, Fan16, Ghi10, Abd10a, Abd10d, Mar11, Ace15, Xia19, 4LAC, Pei16, Pei20SCPMA, Pei20RAA, Fan21, Bur21}, and all of these properties are due to the relativistic beaming effect \citep{MGP87, Ghi93, DG95, Sav10, Fan09, Fan13, Pei19, Yan22}. 

Historically, featured by appearance of the broad emission lines in their optical spectra, blazars are divided into two categories: flat-spectrum radio quasars (FSRQ) that exhibit strong and broad emission lines with equivalent width of EW $\textgreater$ 5 $\dot{A}$ in the rest-frame and BL Lacertae Objects (BL Lacs) that shows quasi-featureless spectra (EW $\textless$ 5 $\dot{A}$) \citep{UP95, SF97}. Alternatively,  the demarcation between two subclasses can be drawn based on the luminosity of the broad-line region (BLR) measured in Eddington units that FSRQs have $L_{\rm BLR}/L_{\rm Edd}\gtrsim5\times10^{-4}$, i.e., $L_{\rm Disk}/L_{\rm Edd}\gtrsim5\times10^{-3}$ when $L_{\rm BLR}\simeq0.1L_{\rm Disk}$ are taken into account, while BL Lacs have less than this criterion \citep{Ghi11, Pei22}, here $L_{\rm BLR}$, $L_{\rm Disk}$ and $L_{\rm Edd}$ denote the broad-line region luminosity, accretion luminosity and Eddington luminosity, respectively. This more physical classification indicates that FSRQs may have stronger accretion disk emission. 

The spectral energy distributions (SEDs) of blazars are usually dominated by two spectral peaks: the first at low to medium energies (radio to X-ray, $\nu^{\rm p}\sim10^{13}$ Hz), and the second at high energies (through X-ray to $\gamma$-ray, $\nu^{\rm p}\sim10^{22}$ Hz) \citep[see e.g.,][and references therein]{Ghi98, Fan16}. Based on the leptonic blazar model, the first peak is produced by synchrotron emission from ultra-relativistic electrons embedded in a magnetic field within the plasma jet, and the second peak is believed to be emanated from the inverse Compton scattering (IC) of low energy photons by the same electron population that generates the synchrotron emission \citep{Fos98, Ghi98, Bot99, Che18}. The GeV emission is generally believed to be generated by IC emission.

Blazars dominate the extragalactic $\gamma$-ray sky \citep{1FGL, 1LAC, 4FGL, 4FGL-DR3}. However, the question of the dominant production mechanism and the exact location of the $\gamma$-ray emission observed in blazars remains unresolved for several decades. The lack of high-resolution instruments and the complicated nature of blazars have resulted in plenty of proposals about the $\gamma$-ray emission region.

Regarding to FSRQs, it is generally accepted that the $\gamma$-ray photons are probably attributable to the IC scattering of external ambient photon fields (EC). Two scenarios have been discussed, namely {\it near-site}, i.e., inside the BLR, which the dissipated energy is located at a distance of $\lesssim0.1-1$ pc from a central supermassive black hole (SMBH) \citep[see e.g.,][]{Sik94, PS10}, and/or {\it far-site}, i.e., outside the BLR or beyond the molecular dusty torus (DT), which the electron energy is dissipated several parsecs away from SMBH \citep[e.g.,][]{Wag95, Bla00, Jor01(a), Arb02, Mar10, Mer19}. 

The critical difference between the BLR and DT is the energy of the seed photons. If the GeV emission originates within the BLR, the IC scattering of ultraviolet (UV) seed photons that produces the $\gamma$-rays occur at the onset of the Klein-Nishima regime. On the other side, if the $\gamma$-ray emission is produced farther out in the DT, then the IC scattering of infrared (IR) seed photons comes about the Thomson regime.   

The Large Area Telescope on board the {\it Fermi} Gamma-ray Space Telescope (Fermi-LAT) is providing for the first time $\gamma$-ray light curves and detecting flares with variability timescales $\sim10^{4}$ s in some FSRQs since its launch in 2008, which offer the evidence of $\gamma$-ray emission produced in the scenario of {\it near-site} \citep{LAT09, 0FGL, 1FGL, Ack10, Abd10b, Fos11, Nal14}. By means of analyzing the light curves of two FSRQs, 3C 454.3 and PKS 1510-089, \citet{Tav10} discussed the implications of significant variability on short timescales have challenged the scenario that $\gamma$-rays are produced in regions of the jet at large distances (tens of parsec) from the black hole. 

The studies of blazars in the radio band or very long baseline interferometry (VLBI) monitoring programs suggest that the $\gamma$-ray and VLBI jet emission are co-spatial, supporting the idea of {\it far-site} scenario \citep[e.g.,][]{Sik08, Lar08, Mar10, Jor10, Agu11}. \citet{Zhe17} studied 36 FSRQs via modeling their SEDs and came to a conclusion that the $\gamma$-ray-emitting regions of FSRQs are located closer to the dusty DT ranges than the BLR. Applying the similar method to SEDs modeling, \citet{Cao13} inferred that the location of the GeV emission region is outside the DT for three quarters of selected FSRQs in their sample. \citet{Lin05} analysed 3C 279, one of the best-observed blazars, proposing that a significant external seed photon field which is provided by DT, extending further than the BLR. Quite recently, \citet{Kra22} derives the de-projected distance between the central engine and the region of the GeV emission for 46 $\gamma$-bright blazars via exploring the correlation between the 15 GHz VLBA flux densities and the $\gamma$-ray photon flux. They ascertained that the seed photons responsible for the $\gamma$-ray emission are likely to originate well beyond the BLR, locating at the distance of a few parsecs from the central engine. 

In this paper, aiming to determine the location of GeV emission for the $\gamma$-ray FSRQs, we follow an effective method firstly proposed by \citet{Geo12}, namely the seed photon factor (SF), and constrain the location of energy dissipation for an enlarged sample of FSRQs. This paper is organized as follows. The approach we apply is going to be well presented in Section \ref{sec2}. In Section \ref{sec3} we describe our sample and the derived results. We conduct the statistical analysis and make discussions in Section \ref{sec4}. Finally, we summarize our main findings in Section \ref{sec5}. Throughout this paper, we adopt the $\Lambda$-CDM model with $\Omega_{\Lambda}\simeq0.73$, $\Omega_{M}\simeq0.27$, and $H_{0}\simeq$ 68 km s$^{-1}$ Mpc$^{-1}$ \citep{Planck14}. The logarithms we employ in all equations below are in base of 10.

\section{METHOD}\label{sec2}

\subsection{Model Description}
The peak energy of synchrotron emission and the EC scattering in the observer frame can be expressed by 
\begin{equation}
\epsilon_{\rm syn}=\displaystyle\frac{B}{B_{\rm cr}}\gamma_{\rm b}^2\delta/(1+z)
\label{eq1}
\end{equation}
\begin{equation}
\epsilon_{\rm EC}=\displaystyle\frac{4}{3}\epsilon_{0}\gamma_{\rm b}^2\delta^2/(1+z)
\label{eq2}
\end{equation}
respectively \citep{CB90, Tav98, GT08}, where $\delta$ is the Doppler factor of the jet, defined by $\delta=[\Gamma(1-\beta\cos\theta)]^{-1}$, $\beta=v/c$ denotes the speed of electrons in units of the speed of light $c$, $\Gamma=(1-\beta^{2})^{-1/2}$ is the bulk Lorentz factor, and $\theta$ signifies the viewing angle. $\gamma_b$ is the Lorentz factor of the electrons responsible for the synchrotron and EC components, $\epsilon_{0}$ is the characteristic energy of the external seed photons, $B$ is the magnetic field permeating the emission region within the jet, and $B_{\rm cr}=m_{e}c^{3}/e\hbar=4.4\times10^{13}$ G is the critical magnetic field. $\epsilon_{\rm syn}$ and $\epsilon_{\rm EC}$ are both in units of electron rest mass energy. 

Taking the ratio of the two peak energies, we obtain
\begin{equation}
\displaystyle\frac{B}{\delta}=\frac{4\epsilon_0\epsilon_{\rm syn} B_{\rm cr}}{3\epsilon_{\rm EC}}.
\label{eq3}
\end{equation}

The observed synchrotron peak luminosity and the observed peak inverse Compton luminosity are given by \citep{BG70, Ryb81}
\begin{equation}
L_{\rm syn}=\frac{4}{3}\sigma_{T}c\beta\gamma_{b}^2 n(\gamma_{b})U_{B}\delta^4
\label{eq4}
\end{equation}
\begin{equation}
L_{\rm IC}=\frac{16}{9}\sigma_{T}c\beta\gamma_{b}^2 n(\gamma_{b})U_{0}\delta^6,
\label{eq5}
\end{equation}
here $\sigma_{T}$ is the Thomson cross section, $n(\gamma_{b})$ is the electron energy distribution at $\gamma_{b}$, $U_{0}$ is the external photon field energy density in the galaxy frame, defined by $U_{0}=\frac{3}{4}U'_{0}/\Gamma^{2}$, where $\Gamma$ is the bulk Lorentz factor, $U'_{0}$ is the external photon field energy density in the jet comoving frame, and $U_B=B^2/8\pi$ is the magnetic field energy density. 

The ratio of Equation (\ref{eq4}) and Equation (\ref{eq5}) is a well-known parameter---Compton dominance (CD),
\begin{equation}
{\rm CD}=\frac{L_{\rm IC}}{L_{\rm syn}}=\frac{32\pi\delta^2\rm U_0}{3B^2}.
\label{eq6}
\end{equation}
  
Substituting Equation (\ref{eq3}) into Equation (\ref{eq6}), one can read 
\begin{equation}
\displaystyle\frac{U_{0}}{\epsilon_{0}^{2}}=\frac{\rm CD}{6\pi}\frac{\epsilon_{\rm EC}^{2}}{\epsilon_{\rm syn}^{2}}B_{\rm cr}^{2}
\label{eq7}
\end{equation}
\begin{equation}
\Rightarrow\displaystyle\frac{\sqrt{U_0}}{\epsilon_0}=10120\times\frac{{\sqrt {\rm CD}}\,\nu_{\rm syn, 13}^{\rm p}}{\nu_{\rm EC, 22}^{\rm p}}\,{\rm G},
\label{eq8}
\end{equation}
where $\nu_{\rm syn, 13}^{\rm p}$ and $\nu_{\rm EC, 22}^{\rm p}$ signify the peak frequency of synchrotron/inverse Compton component in units of $10^{13}$ Hz and $10^{22}$ Hz, respectively. 

Therefore, the seed photon factor (SF) can be attributed as $\log\sqrt{U_0}/\epsilon_0$, i.e.,
\begin{equation}
{\rm SF}=\log\displaystyle\frac{\sqrt{U_0}}{\epsilon_0}=\log\left(10120\times\frac{{\sqrt {\rm CD}}\,\nu_{\rm syn, 13}^{\rm p}}{\nu_{\rm EC, 22}^{\rm p}}\right)\,{\rm G}.
\label{eq9}
\end{equation}  
Here we remark that, $U_o$ and $\epsilon_0$ denotes the energy density and characteristic photon energy of the external seed photon population, respectively, which the former parameter is in units of erg cm$^{3}$, and the latter one is in units of the electron rest mass. Since 1 erg =1 cm$^{2}$ g s$^{-2}$ in cgs units, thus SF $=\log\sqrt{U_0}/\epsilon_0$ will be in units of cm$^{-1/2}$ g$^{1/2}$ s$^{-1}$, i.e., Gauss. This also can be easily told from Equation (\ref{eq7}) that $\sqrt{U_0}/\epsilon_0$ has units consistency with $B_{\rm cr}$ which is in Gauss. Furthermore, this seed photon approach is only applicable to the EC emission model, since our derived estimation of SF is fully based on Equation (\ref{eq2}). Therefore, we only make use of this method on discussing FSRQ-type blazars.

This diagnostic is robust. The value of SF for a specific source can be determined as long as four physical parameters from the broadband SEDs are available. They are the peak frequency and peak luminosity of the synchrotron emission and inverse Compton scattering (the later can be transformed to CD). Note that, these quantities are observable, and can be obtained directly from the quasi-simultaneous multiwavelength SEDs modeling or other reference. Quasi-simultaneous SEDs guarantee that the biases can be minimized due to averaging, and reduce the chance of interband integration mismatches. In other words, the quasi-simultaneous SEDs are unlikely to have X-ray data during a high state while other data is taken during a low state.

The seed factor is believed to owe to the EC scattering on a specific photon population, hinged on the energy density and characteristic photon energy of the upscattered seed photon population. They are known very well for both the BLR and DT. Thus, after the SF value for a source is determined, we can compare it with the canonical SF values of BLR and DT, and thereby constrain the location of $\gamma$-ray emission for this source.  

\subsection{Characteristic Values of Seed Factor for Broad-Line Region and Molecular Torus}

In this paper, for the canonical SF values of BLR, we follow the calculation in \citet{Geo12}. Reverberation mapping of the BLR points to a typical size of $R_{\rm BLR}\approx(1\sim3)\times10^{17}L^{1/2}_{d,45}$ cm \citep{Kas07, Ben09} for AGNs, where $L_{d,45}$ denotes the accretion disk luminosity in units of $10^{45}$ erg s$^{-1}$. The energy density of the BLR can be estimated via $U_{0,\rm BLR}=\xi L_{d}/(4\pi R_{\rm BLR}c)\simeq(0.3-2.6)\times10^{-2}$ erg cm$^{-3}$, here $\xi$ is the BLR covering factor, and we take $\xi=0.1$ in our calculation \citep{GT09}. The value of $U_{0,\rm BLR}$ can be considered hold the same among different sources. Since the BLR SED in the galaxy frame can be approximated by a blackbody with peak frequency of $\nu_{\rm BLR}=1.5\nu_{Ly_{\alpha}}$ \citep{Tav08}, consequently the characteristic photon energy of BLR is $\epsilon_{0,\rm BLR}=3\times10^{-5}$ in units of the electron rest mass. Using these one can obtain the canonical SF values of BLR is SF$_{\rm BLR}=\log\sqrt{U_{0,\rm BLR}}/\epsilon_{0,\rm BLR}\simeq3.26-3.74$ G \citep{Geo12}. A similar result can be referred in \citet{Har20}. 

In the molecular torus, reverberation mapping and near-infrared interferometric studies have been normally effective for radio-quiet sources, e.g., Seyfert galaxies and low luminosity blazars due to their smaller DT region \citep[e.g.,][]{Sug06, Kis11, Poz14}. However, we assume the relation that the radius of the DT scales are proportional to $L^{1/2}_{d,45}$ holds for our sample of FSRQs, i.e., $R_{\rm DT}\approx2.5\times10^{18}L^{1/2}_{d,45}$ cm \citep[see e.g.,][]{GT09, Yan18, Pei22}. The energy density of DT therefore can be estimated with $U_{0,\rm DT}\simeq3.91\times10^{-3}-1.42\times10^{-2}$ erg cm$^{-3}$. The molecular torus spectrum can be approximated using a blackbody spectrum with a temperature of $T=1200$ K \citep{Mal11, Geo12, Har20}, which turns out the characteristic photon energy for the DT is $\epsilon_{0,\rm DT}=5.7\times10^{-7}$. Hence, we obtain the canonical SF values of DT is SF$_{\rm DT}=\log\sqrt{U_{0,\rm DT}}/\epsilon_{0,\rm DT}\simeq5.04-5.32$ G.

%What deserved to be mentioned is that BL Lacs' SF are not accurate extremely using this method, due to the energy of self-synchrotron photos $\epsilon_{\rm SSC}$=$4\epsilon_0\gamma_{\rm b}^2\delta/3(1+z)$ lacks a $\delta$ then barely $\rm B$ acquired not the $\rm B$/$\delta$. However, the articles \citet{Geo12} and \citet{Ada20} use this method to constrain the location of Gamma-ray of BL Lacs, with reasonable results, so we discuss BL Lacs and BCUs with this method as well.

\section{SAMPLE AND RESULTS} \label{sec3}

\subsection{Sample}
To study the black hole mass ($M_{\rm BH}$), accretion luminosity ($L_{\rm disk}$) and other related central engine properties of blazars detected by Fermi-LAT, \citet{Pal21} presented a catalog of 1030 sources for which the broadband SEDs are available. They collected the data from the Space Science Data Center (SSDC) SED builder tool, and also involved the flux measurements given by the Second Swift X-ray Point Source catalog \citep[2SXPS,][]{Eva20} and the 4FGL-DR2 catalog \citep{4FGL}. Then the peak frequencies and corresponding fluxes for the synchrotron and inverse Compton components are estimated by fitting a second-degree polynomial to both peaks using the built-in function provided in the SSDC SED builder tool. Their whole catalog containing the SED fitting results has been made public\footnote{\url{http://www.ucm.es/blazars/engines}}.

However, their origin sample does not provide the classification. We thus cross-check this sample with 4FGL-DR3 \citep{4FGL-DR3} and our previous work \citep{Fan16}. Finally, we collect 572 $\gamma$-ray FSRQs for which the peak frequencies and peak luminosities of synchrotron and EC components are given by \citet{Pal21} (and also the CD values). Besides, we likewise find 103 BCUs, i.e., blazar candidates of uncertain type \citep{3LAC}, in the sample of \citet{Pal21}. For the purpose of enlarging our sample, we classify these sources into UFs (BCUs classified as FSRQs) and UBs (BCUs classified as BL Lacs) by employing an empirical demarcation given by \citet{Che18}, namely, $\alpha^{\rm ph}_{\gamma}=-0.127\log L_{\gamma}+8.18$, where $\alpha^{\rm ph}_{\gamma}$ is the $\gamma$-ray photon index and $L_{\gamma}$ denotes the $\gamma$-ray luminosity in units of erg s$^{-1}$. Based on this criterion, 47 UFs are emerged. Therefore, we overall compile a catalog of 619 FSRQs. We list all the relevant data in Table \ref{tab1}. In this table, column (1) gives the 4FGL name; column (2) the redshift; column (3) the classification, where `FSRQ' denotes the confirmed flat-spectrum radio quasars and `UF' denotes the BCU objects that classified as FSRQ using the criterion from \citet{Che18}; column (4) the peak frequency of synchrotron emission in units of Hz; column (5) the peak frequency of inverse Compton emission in units of Hz; column (6) the peak luminosity of synchrotron component in units of erg s$^{-1}$; column (7) the flux of peak luminosity of inverse Compton component in units of erg s$^{-1}$; column (8) the Compton Dominance; column (9) the reference for col. (4)-(5), where P21=\citet{Pal21}; column (10) the $\gamma$-ray photon index adopted from 4FGL-DR3 \citep{4FGL-DR3}; column (11) the $\gamma$-ray luminosity in units of erg s$^{-1}$ and column (12) the derived SF values in this work ($\log \sqrt{\rm U_0}/\epsilon_0$).   

\subsection{The SF distribution of FSRQs}\label{sec3.1}

\begin{figure}
   \centering
   \includegraphics[width=12cm]{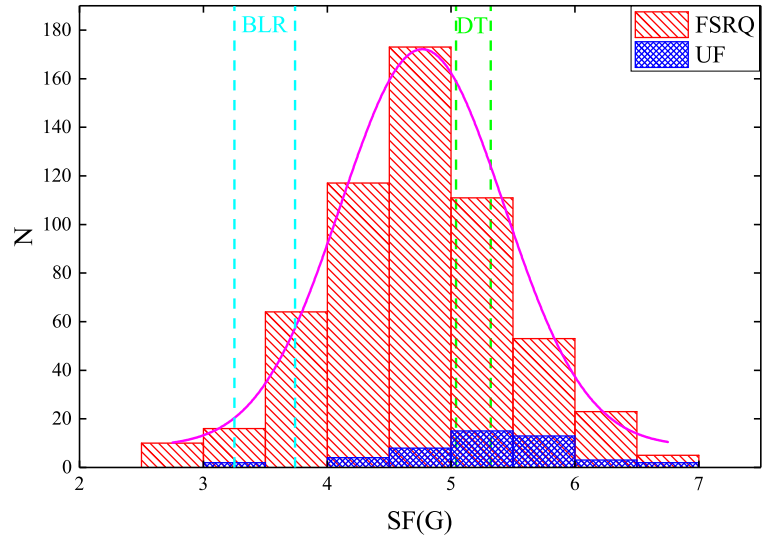}
   \caption{Histogram and Gaussian distribution of derived SF in our sample. UFs denotes the BCUs categorized as FSRQs \citet{Che18}. The dashed blue and green areas signify the location of BLR and DT, respectively.}
     \label{fig1}
\end{figure}

Employing Equation (\ref{eq9}), we obtain the SF values for our sample. We list the derived result in Table \ref{tab1}. The averaged value combining FSRQs and UFs is $<$SF$>=4.76\pm0.03$ G. The histogram of distribution is presented in Figure \ref{fig1}. We label two ranges, dashed blue area indicates BLR and dashed green indicates DT region, respectively. $\mu=4.72\pm0.03$ and $\sigma=0.63\pm0.05$ has been given with performing the Gaussian fitting on total 619 sources. It is noticeable that our derived distribution is mainly located far beyond the BLR and within the DT. There are 570 sources ($>92\%$ of total sample) are located outside the BLR, i.e., SF$>3.26-3.74$ G. Consequently, our finding on the location of $\gamma$-ray emission region supports the {\it far-site} scenario that seed photons are father out at pc scales. Secondly, we also ascertain that the peak of distribution is much closer to DT region than the BLR, which is consistent with the conclusion in \citet{Zhe17}. 

\begin{table}[htp]
\caption{\label{tab1} Sample of Fermi FSRQs}
\centering
\setlength{\tabcolsep}{2.5mm}{
\begin{tabular}{cccccccccccc}
\hline\hline
4FGL Name & Class &  $z$ & $\log \nu^{\rm p}_{\rm syn}$ & $\log \nu^{\rm p}_{\rm IC}$ & $\log L^{\rm p}_{\rm syn}$ & $\log L^{\rm p}_{\rm IC}$ & CD & Ref. & $\alpha^{\rm ph}_{\gamma}$ & $\log L_{\gamma}$ & SF \\
&&&(Hz)&(Hz)&(erg s$^{-1})$&(erg s$^{-1})$&&&&(erg s$^{-1})$&(G)\\
(1) & (2) & (3) & (4) & (5) & (6) & (7) & (8) & (9) & (10) & (11) & (12)\\
\hline

J0001.5+2113	&	FSRQ	&	1.106 	&	13.81	&	20.64	&	43.68	&	45.17	&	30.9	&	P21	&	2.659 	&	46.78	&	6.92	\\
J0004.3+4614	&	FSRQ	&	1.810 	&	12.35	&	21.35	&	43.76	&	44.55	&	6.17	&	P21	&	2.585 	&	46.63	&	4.4	\\
J0010.6+2043	&	FSRQ	&	0.600 	&	12.42	&	22.6	&	42.85	&	43	&	1.41	&	P21	&	2.318 	&	45.21	&	2.9	\\
J0030.6-0212	&	UF	&	1.804 	&	12.59	&	21.48	&	43.94	&	44.99	&	11.22	&	P21	&	2.403 	&	47.41	&	4.64	\\
J0036.9+1832	&	UF	&	1.595 	&	12.7	&	21.96	&	43.53	&	44.42	&	7.76	&	P21	&	2.385 	&	46.52	&	4.19	\\
J0042.2+2319	&	FSRQ	&	1.426 	&	12.27	&	22.35	&	43.67	&	44.39	&	5.25	&	P21	&	2.322 	&	46.56	&	3.29	\\
$\cdots$ & $\cdots$ & $\cdots$ & $\cdots$ & $\cdots$ & $\cdots$ & $\cdots$ & $\cdots$ & $\cdots$ & $\cdots$ & $\cdots$ & $\cdots$\\

\hline
\end{tabular}
}
\tablecomments{Column (1) gives the 4FGL name; column (2) the redshift; column (3) the classification, where `FSRQ' denotes the confirmed flat-spectrum radio quasars and `UF' denotes the BCU objects that classified as FSRQ using the criterion from \citet{Che18}; column (4) the peak frequency of synchrotron emission in units of Hz; column (5) the peak frequency of inverse Compton emission in units of Hz; column (6) the peak luminosity of synchrotron component in units of erg s$^{-1}$; column (7) the flux of peak luminosity of inverse Compton component in units of erg s$^{-1}$; column (8) the Compton Dominance; column (9) the reference for col. (4)-(5), where P21=\citet{Pal21}; column (10) the $\gamma$-ray photon index adopted from 4FGL-DR3 \citep{4FGL-DR3}; column (11) the $\gamma$-ray luminosity in units of erg s$^{-1}$ and column (12) the derived SF values in this work ($\log \sqrt{\rm U_0}/\epsilon_0$).\\
(The table is available in its entirety in machine-readable form)}
\end{table}

\section{DISCUSSION} \label{sec4}

For the past several decades, a variety of discussions have been done on the topic of the emission region of $\gamma$-ray blazars. In the leptonic model, the GeV emission of blazars is produced by IC scattering of photons off of the same relativistic electrons in the jet that contribute to the synchrotron emission. For FSRQs, the seed photons for IC scattering are the synchrotron photons originating external to the jet, e.g., UV photons producing from the BLR, or IR photons producing from the DT. Therefrom, two competing scenarios come forth---{\it near-site} and {\it far-site} scenario, which the former one is based on the idea that the $\gamma$-ray emission region in relativistic jets of blazars is inside the BLR, at $\sim0.1-1$ pc from the central engine, while the later one considers a region much farther away from the central engine at $\gg$1 pc.     

To our knowledge, the distribution of seed photos can reflect the location of $\gamma$-ray emission region, it is believed that seed photons are dissipated within the BLR in the {\it near-site} or dispersed close to the DT in the {\it far-site} scenario. The BLR produces UV photons while the DT produces IR photons.

Constraining the location of $\gamma$-ray emitting region can pave a path for us to better understand the IC radiative mechanism and underlying peculiarity of blazars. If the IC emission dominated by EC process is originated within the BLR, a correlation between $\gamma$-ray and UV flares can be expected since the UV photons embedded in BLR would be available for up-scattering. Alternatively, for a {\it far-site} scenario, the IR photon fields are generated by the DT via reprocessing radiation from the accretion disk or by illumination from the jet synchrotron emission itself, and DT is possible dominant source of seed photons for up-scattering to higher energy \citep{Bre18, AC18}. In this present work, we suggest that the seed photons contribute to the $\gamma$-ray emission are produced in the DT region, thus supporting the {\it far-site} scenario. Many authors also identify this {\it far-site} scenario based on different methods they propose \citep[e.g.,][]{Sik08, Jor10, Zhe17, Kra22}.   

%At the begining, many research discussed the whole blazars, ignoring the classification of which. A variety of results have been porposed. One theory supposed that the Gamma-ray emission of blazars is produced closed to the central engine. \citet{Ghi96} conclude that most of the dissipation is likely to occur as the outflowing material approaches the BLR. The same conclusion is in \citet{Ars18},\citet{Geo01}\citet{Tav2010},\citet{Kov09},\citet{Sik94}. Scenario, there are also many results suppose the other theory, locating in far site(e.g.\citep{Mar18},\citep{Wu18},\citep{Nal14}). \citet{Wu18}  find that the SED fitting with the seed photons from the torus are better than those utilizing BLR photons, which suggests that the ¦Ã-ray emitting region may be located outside the BLR.

%\citet{Jos14} probed into accurate EC-scattered high-energy spectra and consider three sources of seed photon fields, namely the accretion disk, the BLR, and the MT, putting forward that the MT played an important role in contributing toward $\gamma$-ray emission in the MeV range, which fit for our result. 

We consider that, the decisive difference between the BLR and the DT lies in the energy of the seed photons \citep[see e.g.,][]{Dot12}. If the $\gamma$-ray emission originates inside the BLR, the IC scattering of the BLR ultraviolet seed photons producing the $\gamma$-rays takes place at the KN regime. Whereas, If the GeV emission is originated farther out in the pc-scale DT, the IC scattering of the infrared DT seed photons producing the $\gamma$-rays then comes about in the Thomson regime.

\subsection{The dominant location of seed photons} 

In a jet frame, the comoving energy density for an isotropic photon field is $U'\approx(4/3)\Gamma^{2}U$, and $U'\approx(3/4)\Gamma^{-2}U$ for the case that photons coming in the emitting region from behind \citep{DS94}. If the GeV emission is originated within the BLR, the BLR photon field would be regarded as isotropic in the galaxy frame, then $U'_{\rm BLR}\sim(4/3)\times10^{2}\Gamma^{2}$ erg cm$^{-3}$. This consideration that the photon field is isotropic is also suitable for the DT seed photon energy density inside the BLR, i.e., $U'_{\rm DT}\sim(4/3)\Gamma^{2}$ erg cm$^{-3}$. Thus, for the {\it near-site} scenario, a factor $\sim$ 100 is resulting in from the comparison between two above equations. In this sense, the seed photons of BLR ($U'_{\rm BLR}$) are dominant. 

On the other hand, for the {\it far-site} scenario, the BLR photons come in the emission region from behind, then $U'_{\rm BLR}\sim(3/4)\times10^{-8}\Gamma^{-2}$ erg cm$^{-3}$. The DT seed photon energy density remains unchanged. Consequently, if the GeV emission is produced approach to the DT, the DT ($U'_{\rm DT}$) dominates over $U'_{\rm BLR}$.     

\citet{Dot12} suggested that the electron cooling time plays an important role in the difference of the photon energy between BLR and DT, where the $\gamma$-ray-emitting electron IC cooling occurs \citep[see also][]{Cao13}. Specifically, the energy dependence of the electron cooling time can be adopted to determine the regime at which the electrons producing the GeV emission cooling, i.e., the $\gamma$-ray emission takes place whether in the TH regime or KN regime. Our derived result in this paper supports the IC scattering of the IR seed photons take place in the Thomson regime, which is leading to energy dependent electron cooling times, demonstrated as faster cooling times for higher Fermi energy \citep{Dot12, Fin13}

\subsection{The transition from the Thomson to Klein-Nishina regime} 

\begin{figure}[htb]
   \centering
   \includegraphics[width=12cm]{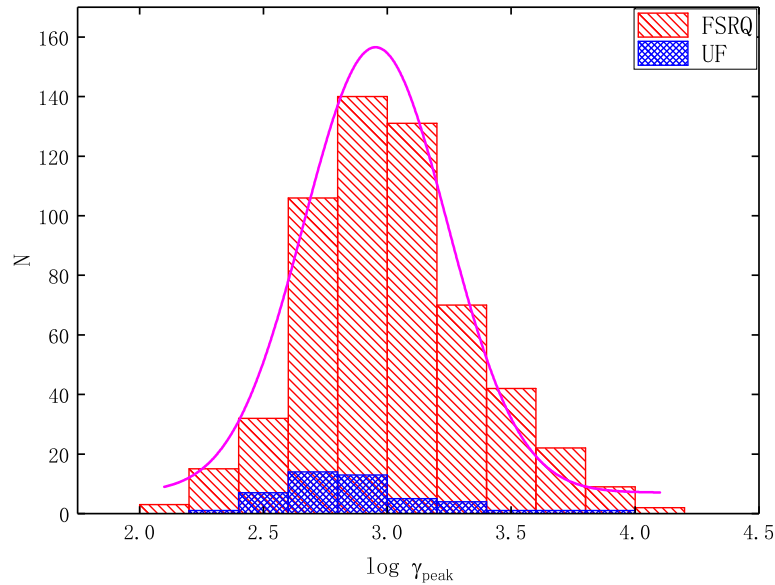}
   \caption{Distribution of the electron Lorentz factor. Gaussian fitting gives $\mu=2.95\pm0.02$ and $\sigma=0.28\pm0.02$. The whole range is under the TH regime.}
     \label{fig2}
\end{figure}

In the TH regime, \citet{Abd10d} presented a tight correlation between the electrons Lorentz factor and the EC peak frequency,
\begin{equation}
\displaystyle\gamma_{\rm peak}^{2}=\frac{3}{4}\frac{\nu^{\rm EC}_{\rm p}}{\nu^{\rm ext}_{\rm p}\Gamma}\frac{1+z}{\delta},
\label{eq10}
\end{equation}  
where $\gamma_{\rm peak}$ denotes the Lorentz factor associated with the jet electrons emitting at the peak of synchrotron emission, $\Gamma$ is the bulk Lorentz factor, and $\nu^{\rm ext}_{\rm p}$ is the peak frequency associated with the external photon filed in the rest frame. We calculate the $\gamma_{\rm peak}$ for our 619 FSRQs, assuming that (i) the Doppler factor of $\delta\approx\Gamma$ for the relativistic jet close to the line of sight in blazars with a viewing angle $\theta<5^{\circ}$ \citep{Jor05, Der15}; (ii) \citet{Che18} shows that the median value of $\delta$ for a large sample of FSRQs is 10.7, thus we perform our calculation by taking $\delta=\Gamma=10.7$; (iii) {\it Spitzer} observations indicates the typical peak frequency of IR dusty DT emission is $\nu^{\rm ext-IR}_{\rm p}\approx3\times10^{13}$ Hz \citep{Cle07, GT09}. Therefore, in the EC scenario, we obtain that $\left<\log\gamma_{\rm peak}\right>=2.99\pm0.34$, ranging from 2.05 to 4.18. We plot the distribution in Figure \ref{fig2}. 

The threshold from TH to KN regime in the EC model is $\gamma_{\rm peak}\Gamma h\nu^{\rm ext}\gtrsim m_{e}c^{2}$, where $h$ is the Planck constant and $m_{e}c^2$ is the electron energy. This criterion translates into $\log\gamma_{\rm peak}\gtrsim5.58$ (note that we adopt $\Gamma=10.7$). Clearly, all sources are in the TH scattering regime, indicating that the EC scattering of the IR seed photons in DT producing the $\gamma$-rays occurs in the TH regime. This result support our finding that the GeV emission for FSRQs is originated around the DT location.

\subsection{Correlations associated with the synchrotron and IC component}

In this work, we compute the seed photons factors for a large sample of FSRQs via employing their SEDs behavior. As a matter of course, we would like to probe some correlations such as the $\gamma$-ray luminosity against the synchrotron peak frequency or IC peak frequency. 

The monochromatic luminosity is given by
\begin{equation}
L=4\pi d^{2}_{\rm L}\nu f_{\nu},
\label{eq11}
\end{equation}
where $f$ is the {\it K}-corrected flux density at the corresponding frequency $\nu$, $d_{\rm L}$ is luminosity distance. The data of $\gamma$-ray emission (e.g., the photon index ${\alpha^{\rm ph}_{\gamma}}$ and $\gamma$-ray flux) are taken from \citet{4FGL-DR3}. Using this equation, we can obtain the $\gamma$-ray luminosities, and list them in Col. (11) of Table \ref{tab1}.

\begin{figure}[htb]
   \centering
   \includegraphics[width=8.9cm]{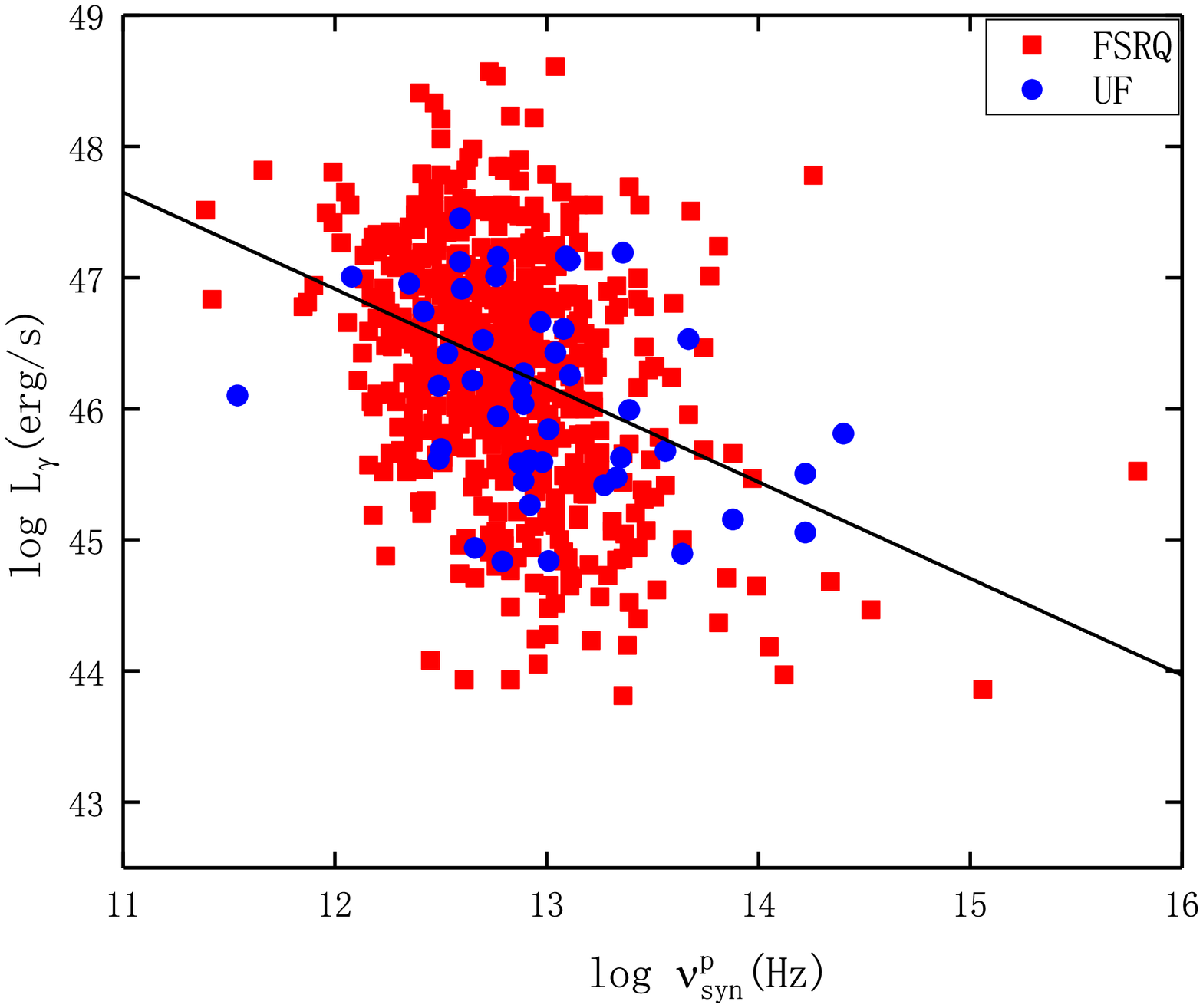}
   \includegraphics[width=8.9cm]{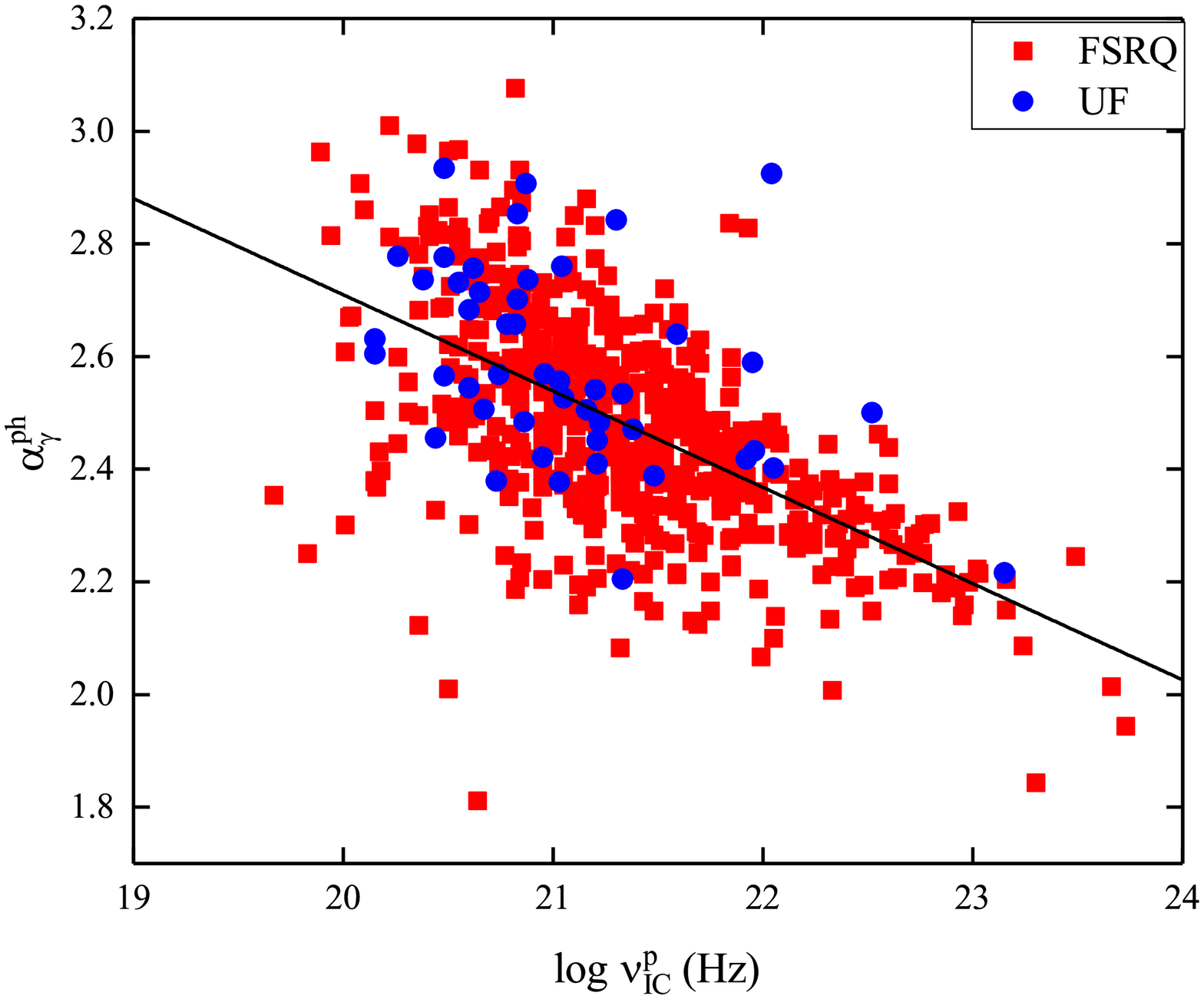}
   \caption{Plots of the $\gamma$-ray luminosity versus the peak frequency of synchrotron emission (left panel) and the photon index against the peak frequency of inverse Compton (right panel). The best-fitting gives $\log L_{\gamma}=-(0.74\pm0.07)\log\nu^{\rm p}_{\rm syn}+(55.74\pm0.96)$ with $r=-0.37$ and $p<10^{-4}$ for the synchrotron component, and $\log\nu_{\rm EC}^{\rm p}=-(2.06\pm0.11)\alpha_{\gamma}^{\rm ph}+(26.44\pm0.28)$ with $r=-0.59$ and $p\sim0$ for the IC component, respectively.}
     \label{fig3}
\end{figure}

The comparisons of $\gamma$-ray luminosity against synchrotron peak frequency and photon index versus IC peak frequency are shown in Figure \ref{fig3}. Two strong correlations are found. The best-fitting for the synchrotron component is $\log L_{\gamma}=-(0.74\pm0.07)\log\nu_{\rm syn}^{\rm p}+(55.74\pm0.96)$ with a correlation coefficient $r=-0.37$ and a chance probability of $p<10^{-4}$ (see the left panel). Similarly, \citet{Fan16} found a tight anti-correlation between $\log\nu_{\rm syn}^{\rm p}$ and $\log L_{\gamma}$ for 999 $\gamma$-ray loud blazars, described as $\log L_{\gamma}=-(0.29\pm0.02)\log\nu_{\rm syn}^{\rm p}+(49.58\pm0.35)$ with $r=-0.32$ and $p<10^{-4}$. \citet{Fos98} also shows the highly significant relation that the synchrotron peak is increasing with decreasing luminosity. The anti-correlation between these two quantities is believed to be due to the strongly beamed in the radio emission. \citet{Nie08} has found that the radio Doppler factor would be larger at low synchrotron peaked blazars and smaller at highly synchrotron peaked blazars. This leads to highly beaming in low peaked sources and lower beaming in highly peaked sources. Thus, an anti-correlation between the peak frequency and radio luminosity can be expected. Meanwhile, the $\gamma$-ray luminosity and radio luminosity for Fermi blazars are strongly correlated \citep[see e.g.,][and references therein]{Fan16, Zha18}. Therefore, an anti-correlation between $\log\nu_{\rm syn}^{\rm p}$ and $\log L_{\gamma}$ can be well explained.    

On the other hand, the right panel demonstrates a significantly negative correlation in the $\alpha_{\gamma}^{\rm ph}-\log\nu_{\rm IC}^{\rm p}$ plane, namely $\log\nu_{\rm EC}^{\rm p}=-(2.06\pm0.11)\alpha_{\gamma}^{\rm ph}+(26.44\pm0.28)$ with $r=-0.59$ and $p\sim0$. Since sometimes there are not sufficient quasi-simultaneous data to construct the IC bump for a large sample, we can make use of this relation to estimate the EC peak frequency when a photon index is given for FSRQs. \citet{Abd10d} also derived the correlation between the IC peak and photon index for 48 sources, which their quasi-simultaneous on GeV band are available to conduct SEDs fitting and obtain the IC peak frequency. The best-fitting is $\log\nu_{\rm IC}^{\rm p}=-4\alpha_{\gamma}^{\rm ph}+31.6$. Notably, the slope of $\alpha_{\gamma}^{\rm ph}-\log\nu_{\rm IC}^{\rm p}$ plane we obtain in this work is $\sim-2$ and $-4$ was found in \citet{Abd10d}. We consider this is owing to the collected sample in \citet{Abd10d} was small, and besides, most principally, combining the FSRQs and BL Lacs. Our precent work only takes FSRQs into account, thus the slope will flatten since the FSRQs normally have a larger $\gamma$-ray photon index.           

Similarly, \citet{AC18} ascertained a tight relation displaying that $\alpha_{\gamma}^{\rm ph}=-0.229\log\nu_{\rm IC}^{\rm p}+7.34$ for a limited sample of radio-Planck sources with good measures for the IC parameter. This tight connection between $\alpha_{\gamma}^{\rm ph}$ and $\log\nu_{\rm IC}^{\rm p}$ indicates that blazars are associated with the steepest $\gamma$-ray sources in the 0.1$-$100 GeV band with the IC peaks around the MeV band. \citet{Ghi09} inferred that reducing the $\gamma$-ray flux threshold may detect blazars with steeper spectral index and lower luminosity \citep[see also][]{Abd09}.

Finally we remark that the correlations between $\gamma$-ray behaviour versus the synchrotron/IC components showing in Figure \ref{fig3} are natural since both components depend straightly on the relativistic electrons within the jet which are producing synchrotron emission and performing for the up-scattering of low-energy photons into $\gamma$-rays \citep[see the detail discussion in][]{Gio12, Gio13}.

\begin{figure}[htb]
   \centering
   \includegraphics[width=12cm]{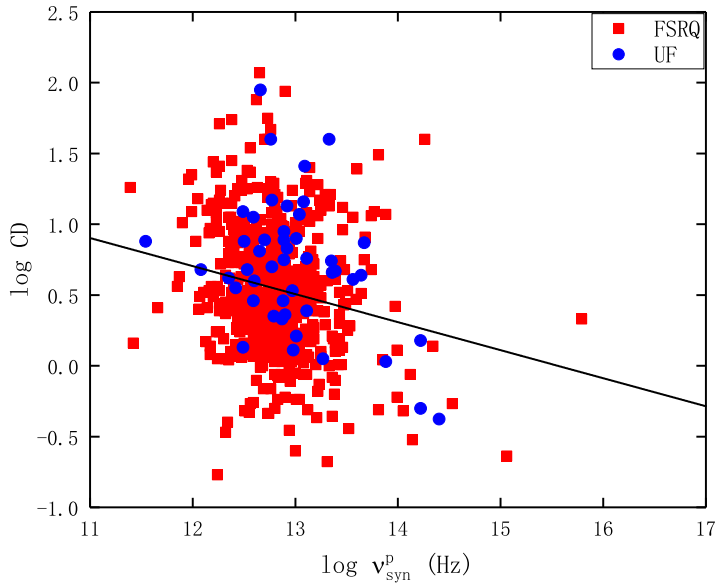}
   \caption{Plot of Compton dominance versus synchrotron peak frequency. The best-fitting gives $\log {\rm CD} =- (0.20\pm0.03)\log\nu^{\rm p}_{\rm syn}+(3.08\pm0.50)$ with $r=-0.20$ and $p=6.32\times10^{-7}$.}
     \label{fig4}
\end{figure}

It should be noted that, the relationships associated the $\gamma$-ray behavior and both the synchrotron and IC contributions are intrinsic, and not depending on the redshift. \citet{Fos98} discovered that the correlations persists even if the redshift effect is subtracted, which elicited the question whether the spectral sequence or `blazar sequence' (an unified scheme whereby blazars continua can be described by a family of analytic curves with the source luminosity as the fundamental parameter) actually exists? \citet{Fin13} probes into the Compton dominance (or named the $\gamma$-ray dominance in the past), the ratio of the peak of the Compton to the synchrotron peak luminosities, which is essentially a redshift-independent quantity and thus crucial to answer this question. Subsequently, they found that a correlation exists between CD and the peak frequency of the synchrotron component for all blazars in their sample, including ones with unknown redshift. In this work, to confirm this verdict, we also analyze the correlation that CD against $\nu^{\rm p}_{\rm syn}$ for our FSRQs sample. An anti-correlation is obtained that described as $\log {\rm CD} =- (0.20\pm0.03)\log\nu^{\rm p}_{\rm syn}+(3.08\pm0.50)$ with $r=-0.20$ and $p=6.32\times10^{-7}$. This plot is shown in Figure \ref{fig4}.

\citet{Yan22} revisits the correlation for a large sample of 255 blazars from 4FGL with available Doppler factors, drawing the conclusion that the observed radio, X-ray, $\gamma$-ray, and synchrotron peak luminosity are all anti-correlated with the peak frequency, but the de-beamed luminosity is positively correlated with the de-beamed peak frequency. This implies that those anti-correlations are perhaps due to a selection effect or a beaming effect \citep[see also][]{Che21}.

\subsection{VHE FSRQs}

Interestingly, \citet{Ach21} investigate 6 bright FSRQs with showing very high energy (VHE) photon emission. The VHE photons are characterized by having an energy $E_{\gamma}\ge20$ GeV in the rest frame of a source, and these FSRQs are also known as TeV sources\footnote{\url {http://tevcat2.uchicago.edu/}}. They found the $\gamma$-ray emission region for these 6 bright FSRQs are beyond the BLR and farther out in the DT, supporting the {\it far-site} scenario. We cross-checked these 6 FSRQs and find they are all in our sample with averaged daily fluxes $10^{-6}$ cm$^{-2}$ s$^{-1}$ within uncertainties of 1$\sigma$ above 100 MeV. They are CTA 102, B2 1520+31, PKS 1510-089, PKS 1502+106, PKS 1424-41 and PKS 0454-234, and their SF values are 4.93, 4.80, 5.15, 4.89, 5.08 and 5.04, respectively. In other words, they are all between the peak of SF distribution and the range of DT, indicating the $\gamma$-ray location of these sources are far beyond the BLR and very close to (or say around) the DT region.

Previous observations of VHE photons also shows that the $\gamma$-ray emission is originated outside the BLR \citep[e.g.,][]{LB06}. However, , \citet{BE16} pointed that the opacity constraints derived can be evaded by resorting to multi-zone models that the production region of GeV and VHE emission is not co-spatially, since the VHE photons observed are emitted at large distances from the SMBH. Therefore, this characteristic feature is likely to challenge the simple one-zone leptonic emission model, which is the standard and most widely used version of the leptonic model assumes there is only one leptonic plasma-filled zone responsible for synchrotron and IC emission of a blazar.

Based on our finding, we suggest that the FSRQs whose their seed photons produced around the DT are  probably TeV sources, showing very high energy ($E_{\gamma}\ge20$ GeV). Another interpretation is that if the emission region is located far beyond the BLR, the VHE photons can avoid the severe $\gamma-\gamma$ absorption and the KN effect would be weak on the VHE spectrum until the energy is far larger than $10^{25}$ Hz \citep[e.g.,][]{Cao13}.

\subsection{Can SF likewise imply the region of neutrino emission?}

On 2017 September 22, the IceCube Neutrino Observatory detected a high-energy ($E_{\nu}\gtrsim290$ TeV) muon-track neutrino event (IceCube-170922A) from the flaring blazar TXS 0506+056 \citep{Ice18(a)}, located at a redshift of $z=0.3365$ \citep{Pai18}. A follow-up analysis of IceCube archival data reported an $\sim3.5\sigma$ excess of $13\pm5$ neutrino events in the range of 32 TeV$-$3.6 PeV to be coincident with the flaring state of the blazar TXS 0506+056 during an $\sim6$ months period in 2014-2015, yielding the first ever $\sim3\sigma$ high-energy neutrino source association \citep{Ice18(b)}. Neutrinos are produced in photopion ($p\gamma$, $p+\gamma\to\pi^{+}+n$) or hadronic ($pp$, $p+p\to X+N_{\pi}\pi^{\pm}$, here $N_{\pi}$ stands for the pion multiplicity) interactions of protons and nuclei.

Beforehand, many authors have proposed that blazars may accelerate protons to very high energy and thus be cosmic neutrino-emitter sources. Notably, flares are ideal periods of neutrino production in blazars \citep[e.g.,][]{Zha20}. During the flares, the density of the target photon field for photomeson interactions with the hadrons within the jet would be enhanced along with the injection rate of accelerated protons. This leads to the neutrino luminosity is significantly reinforced relative to the $\gamma$-ray luminosity, namely $L_{\nu}\propto L^{\alpha}_{\gamma}$, where $\alpha\sim1.5-2$ \citep{Mur14, MW16, Pet16}

Neutrino production is expected to be much more efficient in FSRQs than in BL Lacs due to the higher-powers and existence of external photon fields \citep{AD03}. The nuclear region of FSRQs is naturally abundant in photons, providing an ideal environment to originate high-energy neutrinos from photohadronic interactions. Coincidentally, \citet{Pad19} pointed out that TXS 0506+056 is a masquerading BL Lac object with a hidden BLR of luminosity $\approx5\times10^{43}$ erg s$^{-1}$ and a standard accretion disk, i.e., intrinsically an FSRQ. 

At present, a popular idea for the location of neutrino emission is that the neutrinos and $\gamma$-rays are produced in the same region. The cospatial production is in general expected if the neutrino emission is correlated with the $\gamma$-ray flare, and many studies have reached this conclusion on the 2017 flare of TXS 0506+056 \citep[i.e., a single-zone model, see e.g.,][]{Ans18, Kei18, Cer19, Gao19, Pla20}. Contrastively, \citet{Xue19} proposed a two-zone photohadronic model on the flare of TXS 0506+056 and demonstrates there are two distinct emitting regions, namely a compact region within the BLR responsible for the neutrino emission and $\gamma$-ray emission, and another one region is beyond the BLR account for the synchrotron emission. In this case, large amounts of the neutrinos could be produced than the standard $p\gamma$ model which all the emission is cospatially produced. \citet{Rig20} analyzed the SED of neutrinos and the diffuse flux for FSRQs in two different regions as well. 

Thus, if we consider that the neutrino emission occurs in the same region as the photon emission, there are accordingly two scenarios for describing the location of neutrinos, the {\it near-site} and {\it far-site} scenario, as aforementioned. In consequence, we suggest that the SF can also be an indicator for the location of neutrino-emitting region.

We employ the single-zone model, as the same framework to derive SF. In this model, neutrinos and $\gamma$-rays are co-produced inside the blazar blob through $p\gamma$ interactions. High-energy protons in the emitting region interact with photons to produce charged and neutral pions which leads to comparable fluxes of neutrino emission and $\gamma$-ray emission. If the target photons are comoving with a jet with bulk Lorentz factor $\Gamma$, the predicted neutrino energy $\epsilon_{\nu}$ for all flavors can be estimated via \citep{Oik22}
\begin{equation}
\epsilon_{\nu}\approx100\,\,{\rm PeV}\left(40\,\,{\rm eV}/\epsilon_{\gamma}\right)(\Gamma/10)^{2}(1+z)^{2},
\label{eq12}
\end{equation}
where $\epsilon_{\gamma}$ is the photon energy. Therefore, given the value of external photon field energy density $U_{0}$ for a source, we can obtain the characteristic energy $\epsilon_{0}$ for external seed photons from our derived SF, and compute the photon energy via $\epsilon_{\gamma}=\gamma^{2}_{\rm peak}\epsilon_{0}$.

%3/8ths of proton turns into neutrino production via their energy loosing in $p\gamma$ interactions \citep{Mur14, Rig20},
%\begin{equation}
%\epsilon_{\nu}L_{\epsilon_{\nu}}\approx\displaystyle\frac{3}{8}f_{p\gamma}(\epsilon_{p}L_{\epsilon_{p}})\simeq1.2\times10^{45}\,\,{\rm erg\,\,s}^{-1}\frac{f_{p\gamma}}{10^{-4}}\left(\frac{\epsilon_{p}L_{\epsilon_{p}}}{10^{49.5}\,\,{\rm erg\,\,s}^{-1}}\right)
%\label{eq11}
%\end{equation}
%where $f_{p\gamma}$ denotes the optical depth for protons to $p\gamma$ interactions \citep{Mur16}. The observation of $>$10-100 GeV photons from TXS 0506+056 during the 2017 flare implies that the optical depth for photons to $\gamma\gamma$ interactions on low-energy photons with $\tau_{\gamma\gamma}<1$, which constrains the upper limit to $f_{p\gamma}<10^{-3}$.

%The proton energy can associate with photon energy by
%\begin{equation}
%\epsilon_{p}=6\,\,{\rm PeV}\left(\displaystyle\frac{\epsilon_{\gamma}}{15\,\,{\rm GeV}}\right)
%\label{eq12}
%\end{equation}
%Therefore, we can obtain the characteristic energy $\epsilon_{\gamma}=\epsilon_{0}$ for external seed photons from our derived SF, by adopting a given value of external photon field energy density $U_{0}$, and then compute the proton energy.

%Finally, the proton and neutrino luminosity can be determined by $\epsilon_{\rm i}L_{\epsilon_{\rm i}}=\int_{\epsilon_{\rm i}}L_{\epsilon_{\rm i}}d\epsilon_{\rm i}$, where {\rm i} denotes $p$ and $\nu$.  

\begin{figure}[htb]
   \centering
   \includegraphics[width=12cm]{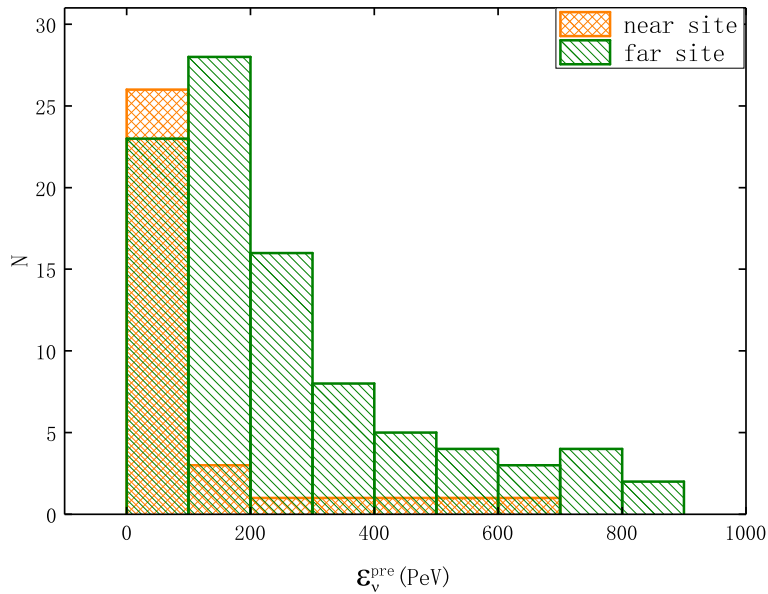}
   \caption{Distributions of predicted neutrino energy in all flavors for {\it near-site} and {\it far-site} scenario, respectively.}
     \label{fig5}
\end{figure}

Our present work manifests that the seed photons are probably originated from the DT region. The SF values span from 5.04 to 5.32 G, and 93 sources in our sample are located in this range. If we consider the neutrino emission is also produced from DT, i.e., a {\it far-site} scenario, we can obtain the neutrino energy is $\epsilon_{\nu}\simeq5.62$ to 850.33 PeV with a median value of 193.20, using $\epsilon_{0,\rm DT}=5.7\times10^{-7}$ and Equation (\ref{eq12}). The distribution of $\epsilon_{\nu}$ for {\it far-site} scenario is displayed in Figure \ref{fig5}.   

In comparison with the PeV neutrino production, $\sim0.1-1$ EeV neutrinos are produced by way of interactions between protons and IR photons from the dust torus. The lower limit of neutrino energy in this circumstances can be roughly estimated by \citep{Mur14}
\begin{equation}
\epsilon_{\nu}=0.066\,\,{\rm EeV}\left(T_{\rm IR}/500\,\,{\rm K}\right)^{-1},
\label{eq13}
\end{equation}
where $T_{\rm IR}$ refers to the aforementioned temperature of the dust torus. In like manner, we adopt $T_{\rm IR}=1200$. Hence, we ascertain $\epsilon_{\nu}\gtrsim0.0275$ EeV.

By contrast, if we consider the {\it near-site} scenario, e.g., SF $=3.26-3.74$ G and there are 34 FSRQs in this region. The characteristic photon energy for BLR is $\epsilon_{0,\rm BLR}=3\times10^{-5}$, which resulting in $\epsilon_{\nu}\simeq2.24$ to 604.34 PeV with a median value of 60.47 (see the distribution in Figure \ref{fig5}). 

The neutrino energy we achieve is typically higher than 1 PeV and the Glashow resonance energy at 6.3 PeV (for electron antineutrinos). We conclude that the external radiation field plays an important role in PeV$-$EeV neutrino production. The predicted $\epsilon_{\nu}$ distributions for the {\it near-site} and {\it far-site} scenario we propose in this work that corresponding to the $\gamma$-ray emission region, are comparable. Thus these two frameworks could be both potential for the production of neutrinos. This remains to be tested during long-lived blazars flares. 

All in all, blazars are the most well-known class of extragalactic high-energy particle accelerators. Especially for FSRQs, since they possess an efficient accretion disk and a dense ultraviolet photon field surrounding the black hole, which are optimal for the production of $\gg$ PeV-energy neutrinos. We propose the seed photon factors can also be an indicator for the neutrino emission location since the idea that neutrinos and $\gamma$-rays are co-produced in a same region is generally accepted.

%\begin{eqnarray}
%\frac{\partial\tau_{\gamma\gamma}}{\partial\Phi}|_{M}&=&\left[22.5\times\Phi^{1.5}(1-\cos\Phi)-9\times\frac{2\alpha_{X}+3}{2\alpha_{\gamma}+8}\Phi^{2.5}\sin\Phi-\frac{2\alpha_{X}+3}{2\alpha_{\gamma}+4}kM_{7}^{-1}A^{-\frac{2\alpha_{X}+3}{2}}(1-\cos\Phi)^{-\frac{2\alpha_{X}+3}{2\alpha_{\gamma}+8}}\sin\Phi\right] \nonumber \\
%&\times&\left[(1-\cos\Phi)^{-\frac{2\alpha_{X}+2\alpha_{\gamma}+11}{2\alpha_{\gamma}+8}}A^{-\frac{2\alpha_{X}+3}{2}}\right]=0,
%\label{eq}
%\end{eqnarray}

\section{CONCLUSION}\label{sec5}

In this paper, we constrain the dissipative location of $\gamma$-ray emission for a large sample of 619 FSRQs, including 572 confirmed FSRQs and 47 BCU-FSRQs candidates, by means of the seed photon factor approach which proposed by \citet{Geo12}. This method is tightly associated with the SEDs behavior, and the SF can be simply derived as long as the peak frequency and peak luminosity of the synchrotron along with inverse Compton components are available. We take these SED data directly from \citet{Pal21}. We stress that this seed photon approach is only valid for the EC scattering model, thus it is applicable to discuss the GeV emission of FSRQ-type blazars. We find that the location of $\gamma$-rays is far beyond the BLR and close to the DT, which supports the {\it far-site} scenario. Meanwhile, we discuss the correlations associated with the synchrotron emission and EC scattering. We also shed new light on the production region of neutrino emission. The main conclusions of this work are as follows:

\begin{enumerate}
\item The region of $\gamma$-ray emission of FSRQs is father out at pc scales from the SMBH where the molecular torus IR emission dominates, which can be illustrated by two aspects: (i) The SF distribution is mainly located far beyond the BLR and verge on the DT region. (ii) All sources are in the Thomson scattering regime, indicating that the EC scattering of IR seed photons in DT producing the $\gamma$-rays occurs in the TH regime, which evidences the GeV emission for FSRQs is originated around the DT region. Besides, the DT dominance also supports that energy dependent electron cooling times, demonstrating as faster cooling times for higher Fermi energy.
\item FSRQs with their seed photons producing in the range of DT are probably TeV sources, which emit very high energy with $E_{\gamma}\ge20$ GeV (VHE). The reason is that if the emission region is located far beyond the BLR, the VHE photons can avoid the severe $\gamma-\gamma$ absorption and the KN effect would be weak on the VHE spectrum until the energy is far larger than $10^{25}$ Hz.
\item An expected anti-correlation between the synchrotron peak frequency and $\gamma$-ray luminosity has been verified for our sample, since the $\gamma$-ray luminosity and radio luminosity for Fermi-detected blazars are strongly correlated, leading to highly beaming in low peaked sources and lower beaming in highly peaked sources.
\item We suggest that our derived SF can also be an indicator to estimate the location of neutrino emission. We also propose that two similar scenarios could be discussed---{\it near-site} and {\it far-site} model---in consideration of the neutrinos and $\gamma$-rays are cospatially produced within a same region. We compute the predicted neutrino energy for overall sample and find $\epsilon_{\nu}$ for two frameworks are comparable with $\gg$ 1 PeV, satisfying our general expectation that blazars are the optimal production factories for PeV-EeV energy neutrinos in the extragalactic $\gamma$-ray sky.   
\end{enumerate}

\begin{acknowledgements}
We are grateful to the anonymous referee for valuable comments and constructive suggestions, which helped us to improve the manuscript. This work is partially supported by the National Natural Science Foundation of China (NSFC U2031201, NSFC 11733001, U2031112), Guangdong Major Project of Basic and Applied Basic Research (Grant No. 2019B030302001). D.Y. Huang acknowledges support from Guangzhou University and ``Challenge Cup" National Undergraduate Curricular Academic Science and Technology Works Competition (Grant 2022TZBNAI2001). Z.Y. Pei acknowledges support from National Science Foundation for Young Scientists of China (Grant 12103012), and China Postdoctoral Science Foundation (Grant 2022M710868). We also acknowledge the science research grants from the China Manned Space Project with NO. CMS-CSST-2021-A06, and the supports for Astrophysics Key Subjects of Guangdong Province and Guangzhou City. 

\end{acknowledgements}

\bibliography{Huang-2022}{}
\bibliographystyle{aasjournal}

\end{document}